\documentclass[aps,showpacs]{revtex4}
\usepackage{graphicx}
\newcommand{\be}{\begin{eqnarray}}
\newcommand{\ee}{\end{eqnarray}}
\newcommand{\no}{\nonumber}
\newcommand{\ba}{\begin{array}}
\newcommand{\ea}{\end{array}}
\newcommand{\tr}{\mbox{Tr}\,}
\newcommand{\str}{\mbox{str}\,}
\newcommand{\e}{\epsilon}
\newcommand{\Si}{\Sigma}
\newcommand{\D}{\Delta}

\begin{document}
\title{Long range disorder and Anderson transition
 in systems with chiral symmetry}
\author{Antonio M. Garc\'{\i}a-Garc\'{\i}a}
\affiliation{Laboratoire de Physique Th\'eorique et Mod\`eles 
 Statistiques, B\^at. 100, \\ Universit\'e de Paris-Sud, 
 91405 Orsay Cedex, France}
\author{Kazutaka Takahashi}
\affiliation{Theoretische Physik III, 
 Ruhr-Universit\"at Bochum, 44780 Bochum, Germany}
\date{\today}

\begin{abstract}
 We study the spectral properties of a chiral random banded matrix (chRBM) 
 with elements decaying as a power-law 
 ${{\cal H}_{ij}}\sim |i-j|^{-\alpha}$. 
 This model is equivalent to a chiral 1D Anderson Hamiltonian 
 with long range power-law hopping.   
 In the weak disorder limit we obtain explicit nonperturbative
 analytical results for the density of states (DoS) and 
 the two-level correlation function (TLCF)
 by mapping the chRBM onto a nonlinear $\sigma$ model. 
 We also put forward, by exploiting the relation between the chRBM
 at $\alpha=1$ and a generalized chiral random matrix model,  
 an exact expression for the above correlation functions.
 We give compelling analytical and numerical evidence that 
 for this value the chRBM reproduces all the features of 
 an Anderson transition. 
 Finally we discuss possible applications of our results 
 to quantum chromodynamics (QCD).
\end{abstract}
\maketitle

\section{Introduction}


 The spectral properties of a disordered system, 
 namely, a noninteracting particle in a random potential, 
 are strongly affected by the global symmetries of the Hamiltonian.
 In the ergodic/metallic regime, the spectral correlations 
 of generic quantum complex systems
 with translational invariant spectrum fall 
 in three universality classes ('standard' from now on) 
 attending to time reversal and spin symmetries. 
 The simplest representative of each universality class corresponds to 
 an ensemble of matrices with the appropriate symmetry 
 and random Gaussian entries. 
 Typical properties of the spectral fluctuations of these matrices 
 include level repulsion and spectral rigidity. 
 In the literature they are usually referred as 
 Wigner-Dyson (WD) statistics \cite{WD}.
 

 In recent years new ('nonstandard' from now on) universality classes 
 have been put forward by relaxing 
 the translational invariance condition \cite{SV,SN,zirn,AZ}.
 These classes are related to additional symmetries of the system. 
 For instance, for (random) matrices with block structure
\be
 \label{rb}
 \mathcal{H} = \left(\ba{cc} 0 & C \\ C^\dag & 0 \ea\right),
\ee
 where $C$ is a general $n\times (n+\nu)$ real, complex or quaternionic matrix, 
 the eigenvalues come in pairs of $\pm \epsilon_i$ or are zero.
 This discrete symmetry is usually called chiral and 
 induces an additional level repulsion around zero 
 which results in different spectral correlations for eigenvalues 
 near zero (the origin) and away from zero (the bulk).
 In the bulk the spectral correlations are not affected by 
 the block structure and WD statistics applies. 


 The motivation to study these nonstandard symmetry classes 
 comes from different branches of theoretical physics. 
 In the context of QCD, the Dirac operator 
 in a chiral basis has a similar block structure. 
 It turns out that in the infrared (ergodic) limit,  
 the eigenvalue correlations of this operator
 do not depend on the dynamics details of the QCD Lagrangian
 but only on the global (chiral) symmetries 
 of the QCD partition function \cite{SV}. 
 Thus random matrices with the correct chiral symmetry of QCD 
 (termed chiral random matrices) \cite{V} accurately describe 
 the spectral properties of the QCD Dirac operator up to some scale 
 known as the Thouless energy \cite{VO}. 
 As in the standard case, depending on the invariance 
 under time reversal symmetry and spin, 
 there exist three chiral random matrix ensembles. 
 In the context of QCD each universality class is related to 
 both the number of colors and the fermionic representation considered.


 Another interesting problem that falls beyond the standard classification
 is that of the spectral correlations of a mesoscopic disordered system 
 composed of both normal conducting and superconducting parts. 
 In this case, in an appropriate basis, 
 the mean field Bogouliobov-de Gennes Hamiltonian can be expressed as
\be
 {\cal H} = \left(\ba{cc} h & \D \\ -\D^* & -h^T \ea\right)
\ee
 where $\Delta$ is a matrix representing the pairing field 
 and $h$ is a matrix representation of the disordered free Hamiltonian. 

 The Hamiltonian $\cal H$ can be effectively modeled as a random matrix
 provided that the phase shift due to reflection 
 in the NS interphase vanishes on average \cite{AZ}. 
 In this situation the gap at the chemical potential 
 disappears and thus pseudo particle excitations
 appear at arbitrarily low energies. 
 The resulting spectral properties also fall apart from WD statistics.
 In this case four new universality classes
 are found attending to spin and time reversal symmetries. 
 Other systems with similar nonstandard symmetries include 
 random flux models \cite{rfm} (here disorder is placed not 
 in the site but in the link (bond disorder)) 
 and bosons in random media \cite{GC}.


 As a general rule (valid for all standard and nonstandard symmetry classes), 
 localization effects tend to gradually erase the impact of symmetries.       
 However, the transition to localization is not the same in both cases.  
 In systems with nonstandard symmetries the localization 
 properties are in general sensitive to details as
 the microscopic form of the random potential \cite{dwv} 
 or the number (odd or even) of lattice points considered \cite{wire}. 
 In certain situations \cite{gade,PL,furu}, for energies close to zero, 
 the DoS diverges and eigenvectors remain delocalized 
 even in the strong disorder regime in 1D or 2D.
 By contrast, in the standard case the spectral correlations  
 are insensitive to the microscopic details of the dynamics provided
 that the disordered potential is short-range.
 The eigenstates of a free particle in a disordered medium in less than 3D 
 are localized in the thermodynamic limit for any amount of disorder.   
 In three and higher dimensions there exists a metal insulator 
 transition (Anderson transition) for a critical amount of disorder. 


 For finite systems, the dimensionless conductance $g=E_c/\D$ 
 ($E_c=\hbar/t_c$ is the Thouless energy, 
 $t_c$ is the classical time to cross the sample diffusively, 
 and $\D$ is the quantum mean level spacing) is a useful magnitude 
 to quantify the deviations from WD statistics 
 due to wavefunction localization. 
 We recall that WD statistics applies 
 in the metallic/delocalized regime $g\to\infty$.
 Nonperturbative corrections due to a finite $g\gg 1$ (weak disorder) 
 were recently evaluated by Andreev and Altshuler \cite{AA}
 by mapping the localization problem onto 
 a supersymmetric nonlinear $\sigma$ model \cite{efetov}. 
 They managed to express the TLCF 
 in terms of the spectral determinant of the classical diffusion operator.   
 As disorder strength increases, localization becomes dominant 
 $g \sim 1$ and eventually the system undergoes a metal insulator transition. 
 Unfortunately the above analytical tools 
 cease to be applicable in this region.
 Numerical simulations of 3D short range Anderson models suggest that, 
 at the Anderson transition, the wavefunction moments $P_q$ present 
 anomalous scaling with respect to the sample size $L$, 
 $P_q=\int d^dr |\psi({\bf r})|^{2q}\propto L^{-D_q(q-1)}$,
 where $D_q$ is a set of exponents describing the transition \cite{wegner}. 
 Wavefunctions with such a nontrivial scaling are said to be multifractal 
 (for a review see Ref.\cite{janssen}). 
 Spectral fluctuations at the Anderson transition 
 (commonly referred as 'critical statistics' \cite{KMu}) 
 are intermediate between WD and Poisson statistics. 
 Typical features include: scale invariant spectrum \cite{SSSLS},
 level repulsion, and sub-Poisson number variance \cite{chi}. 
 Different generalized random matrix model (gRMM) have been successfully 
 employed to describe critical statistics \cite{gRMM,prbm}.


 The study of these gRMM have shown that critical statistics and 
 multifractal wavefunctions can be reproduced in the weak coupling regime 
 $g \gg 1$ by allowing long range hopping in the original Anderson model. 
 Actually power-law long range hopping, far from being a mathematical curiosity, 
 appears in a broad range of systems: 
 glasses with strong dipole interactions in real space \cite{levitov}, 
 the evolution operator of Floquet systems \cite{AL} 
 with a nonanalytical potential 
 or in Hamiltonians leading to classical anomalous diffusion \cite{ant5}. 
 Although the introduction of long range hopping dates back to 
 the famous Anderson's paper on localization \cite{anderson}, 
 these models did not attract too much attention 
 until the numerical work of Oono \cite{YO,YU} and 
 the renormalization group (RG) treatment of Levitov \cite{levitov} 
 in the context of glassy systems. 
 The main conclusion of these works was that power-law hopping 
 may induce an Anderson transition in any dimension provided
 that the exponent of the hopping decay matches the dimension of the space.
 

 A 1D version of this problem, a RBM with a power-law decay,
 was discussed in Ref.\cite{prbm}. 
 By mapping the problem onto a nonlinear $\sigma$ model 
 with a nonlocal interaction, it was shown that  
 for a $1/r$ band decay the eigenstates are multifractal 
 and the spectral correlations resemble the ones at the Anderson transition.
 In this case the TLCF in the $g \gg 1$ limit 
 is expressed in terms of the spectral determinant 
 of a classical anomalous diffusion operator related to 
 classical ballistic diffusion and $1/f$ noise \cite{MK}. 
 Recent investigations \cite{prbm2} have further 
 corroborated the close relation between this random banded matrix model 
 and the Anderson model at the metal insulator transition. 


 In this work we want to study the interplay 
 between long range disorder and chiral symmetry. 
 Our main aim is to provide a detailed analytical account on 
 how the transition to localization occurs in systems with chiral symmetry. 


 The organization of the paper is as follows. 
 In Sec.\ref{chRBM}, 
 we introduce a chRBM with power-law decay 
 (equivalent to a 1D Anderson model
 with power-law hopping and chiral symmetry).
 After mapping it onto a nonlinear $\sigma$ model 
 we investigate the localization properties 
 by using the renormalization group formalism.
 In Sec.\ref{level}, 
 analytical results for both the DoS and the TLCF 
 are obtained in different domains.
 We show for a special value of the power-law exponent ($\alpha=1$)
 the resulting expression is greatly simplified.
 In Sec.\ref{comp}, 
 by exploiting the relation between the chRBM for $\alpha=1$ 
 and an exactly solvable random matrix model, 
 we give an exact result 
 for the DoS and the TLCF valid in the weak disorder regime. 
 It is shown that the spectral correlations for this 
 special value resemble 
 those of a disordered system at the Anderson transition.
 We also conjecture an expression for 
 the DoS and the TLCF  
 for arbitrary power-law exponent.
 These analytical findings are supported by numerical results 
 from direct diagonalization of the chRBM.
 Finally, in Sec.\ref{app}, 
 we discuss applications of our model in the context 
 of QCD \cite{SV}.

\section{The chiral RBM with power-law disorder}
\label{chRBM}

 We study the spectral properties of an ensemble of
 Hermitian chRBM given by 
\be
 {\cal H} = \left(\ba{cc} \omega_1 & \omega_2 \\ 
 \omega_2 & -\omega_1 \ea\right), \label{rb2}
\ee
 where $\omega_{1,2}$ are $n\times n$ Hermitian matrices.
 This form is related to Eq.(\ref{rb}) at $\nu=0$ 
 by a unitary transformation.
 In this paper we restrict ourselves to the $\nu=0$ case. 
 The matrix elements $(\omega_{1,2})_{ij}$ are independently distributed
 complex Gaussian variables with zero mean and variance
\be
 \label{corr}
 \langle |(\omega_{1,2})_{ij}|^2\rangle 
 = a^2(r)
 = \frac{1}{1+\left(r/r_0\right)^{2\alpha}},
\ee
 where $r_0$ and $\alpha$ are real parameters and 
 $r=|i-j|$ is the spectral distance. 
 The choice of complex matrix elements corresponds to 
 a matrix model with unitary symmetry.
 Due to the chiral symmetry, the eigenvalues of Eq.(\ref{rb2}) come 
 in pairs of $\pm \epsilon_i$. 
 This feature induces thus an additional level repulsion around zero 
 and consequently the spectral correlations 
 for eigenvalues near zero (the origin)
 and away from zero (the bulk) are essentially different.
 At the bulk the spectral correlations are not affected by the block 
 structure and coincide with the nonchiral version of Eq.(\ref{rb2}) which
 has been intensively studied in recent years \cite{prbm,prbm2}.

 In this section we study the localization properties at the origin 
 by mapping the chRBM (\ref{rb2}) 
 onto a supersymmetric nonlinear $\sigma$ model \cite{efetov}.
 The above chRBM can also be interpreted as a 1D chiral Anderson model 
 with long range disorder. 
 In this context the parameter $r_0$ measures 
 the amount of off-diagonal disorder.
 Thus the $r_0 \gg 1$ ($r_0 \ll 1$) limit corresponds with 
 the weak (strong) disorder regime.
 In this paper, due to technical reasons 
 (large-$r_0$ allows us to use the saddle-point 
 approximation in the derivation of the nonlinear $\sigma$ model), 
 we are focused on the weak disorder limit.
 Concerning the power-law exponent $\alpha$, 
 we are mainly interested in the range $1/2<\alpha \leq 3/2$. 
 It is known that the spectral correlations
 for the nonchiral RBM case 
 $\alpha\leq 1/2$ are well described by WD statistics \cite{prbm}. 
 We expect this holds also in the present chiral case 
 and we do not consider it in this paper. 
 On the other hand, as shown below, for $\alpha > 3/2$
 our model is similar to a standard 1D short range chiral Anderson model 
 recently investigated in Ref.\cite{ktaka}.  

\subsection{Supersymmetric nonlinear $\sigma$ model}

 We use the supersymmetry method \cite{efetov} to derive the sigma model.
 We shall follow Efetov's notations and conventions.
 As usual, the first step is to express
 the correlation function as a ratio of two determinant. 
 For the one point retarded Green function,
\be
 G^{(R)}(\epsilon) = \frac{1}{\e^{+}-{\cal H}}
 = \frac{1}{2}\left.\frac{\partial}{\partial J}Z[J]\right|_{J=0},
\ee
 where $\e^+=\e+i\delta$ is the energy with small imaginary part
 and the generating function $Z[J]$ is given by 
\be
 Z[J] = \frac{\mbox{det}(\e^{+}-{\cal H}+J)}
 {\mbox{det}(\e^{+}-{\cal H}-J)}.
\ee
 The source field $J_{ij}$ is a $2n\times 2n$ complex matrix.
 The above ratio can be written as a Gaussian integral over 
 a supervector with both fermionic and bosonic components as 
\be
 Z[J] = \int {\cal D}(\psi,\bar{\psi})\exp
 \left[i\bar{\psi}(\e^{+}-{\cal H}+kJ)\psi\right],
\ee 
 where $k=\mbox{diag}(1,-1)$ in superspace, 
 $\psi$ is a $4n$-component supervector and $\bar{\psi}=\psi^\dag$. 

 Following the standard method, 
 we perform the spectral averaging over $\cal H$ and 
 introduce the 4$\times$4 supermatrix $Q$
 using the Hubbard-Stratonovich transformation. 
 After these steps are carried out, the generating function is given by
\be
 \left<Z[J]\right> = \int {\cal D}Q\exp\biggl[
 -\frac{A_0}{2}\sum_{i,j=1}^{n}(A^{-1})_{ij}\str Q_iQ_j 
 +\str\ln
 \left(\e^{+}\Si_z+kJ\Sigma_z+i\sqrt{A_0}Q\right)
 \biggr],
\ee
 where $A_{ij}=a^2(|i-j|)$, $A_0=n^{-1}\sum_{i,j=1}^nA_{ij}$,  
 $\Sigma_z$ is the Pauli matrix in chiral space and 
 the $Q$-matrix verifies $\{Q,\Sigma_x\}=0$.
 In the limit $A_0\sim r_0\gg 1$ the integral over $Q$ 
 can be performed by the saddle-point method. 
 It is not hard to show that in this case the solution of 
 the saddle point equation is $Q_i=\Sigma_z$ at $\e=0$. 
 Consequently the saddle point manifold $Q^2=1$ is
 parametrized as $Q=T\Sigma_z\bar{T}$ where 
 $T$ is the matrix with symmetry specified below and $\bar{T}=T^{-1}$.
 In addition to that, the massive mode changing the saddle point 
 should also be integrated out.
 Unlike the standard case, the effect of these modes plays 
 an important role in our case.
 The matrix $Q$ is thus parametrized as 
\be
 Q_i = T_i(\Si_z+\delta Q_i)\bar{T}_i, \quad 
 \delta Q_i = \delta q_i\Sigma_z, \label{deltaQ}
\ee
 where $\delta q_i$ is a 2$\times$2 supermatrix.
 In order to proceed we first perform an expansion in powers of 
 $\delta Q$ up to second order, and then integrate over it 
 (it is just a Gaussian integration).
 After these manipulations we find
 $\left<Z[J]\right>=\int{\cal D}Q\exp(-F)$ with 
\be
 F = \frac{A_0}{2}\sum_{ij}R_{ij}\str Q_iQ_j
 -\frac{1}{4}\sum_{ij}R_{ij}
 \str \bar{T}_j T_i\Si_x  \str \bar{T}_i T_j\Si_x
 +\frac{i\pi\e}{2n\Delta}\sum_i\str \Si_zQ_i 
 +\frac{i\pi}{2n\Delta}\sum_{ij}\str J_{ij}k\Si_zQ_j, \label{F0}
\ee
 where $R_{ij}=(A^{-1})_{ij}-n^{-1}\delta_{ij}\sum_{kl}(A^{-1})_{kl}$,
 and $\Delta=\pi\sqrt{A_0}/2n$ is the inverse of the DoS 
 (mean level spacing) at $\e=0$.

 We remark that the above results can be applied to any $a(r)$
 provided that $A_0$ is large enough.
 For technical details we refer to Ref.\cite{ktaka}
 where a $a(r)$ with exponential decay was considered.
 Compared with the sigma model for the nonchiral RBM, 
 we found an additional contribution having the double-supertrace form 
 (second term of Eq.(\ref{F0})).
 This term was derived by integrating over the massive modes. 
 We recall that in its derivation  
 it was crucial that the massive modes were parametrized as in Eq.(\ref{deltaQ}). 

 In order to proceed we have to make a gradient expansion of 
 the kinetic term in Eq.(\ref{F0}). 
 As in the standard case \cite{prbm}, 
 this expansion depends on the value of $\alpha$. 
 For $\alpha > 3/2$ 
\be
 \frac{A_0}{2}\sum_{ij}R_{ij}\str Q_iQ_j
 &\sim& -\frac{A_0}{4}
 \sum_{ij} (i-j)^2R_{ij}\str (\nabla Q_i)^2 \no\\
 &\sim& -\frac{A_0^{(2)}}{4A_0}
 \int dx\str (\nabla Q(x))^2,
\ee
 where
\be
 & & A_0
 = 2\int_{0}^\infty dr\frac{1}{1+(r/r_0)^{2\alpha}}
 =  2r_0 \frac{\frac{\pi}{2\alpha}}{\sin\frac{\pi}{2\alpha}}, \no\\
 & & -A_0^{(2)}
 = 2\int_{0}^\infty dr\frac{r^2}{1+(r/r_0)^{2\alpha}} 
 = 2r_0^3 \frac{\frac{\pi}{2\alpha}}{\sin\frac{3\pi}{2\alpha}}, \no\\
 & & -\frac{A_0^{(2)}}{A_0} 
 =  r_0^2\frac{\sin\frac{\pi}{2\alpha}}{\sin\frac{3\pi}{2\alpha}}.
\ee
 We see that the integral in $A_0^{(2)}$ is well-defined for $\alpha> 3/2$.
 After a similar expansion in the double trace term above,
 we find the normal diffusive $\sigma$ model for $\alpha> 3/2$ 
\be
 F =
 \frac{i\pi\e}{2\D V}\int dx\str \Sigma_zQ(x)
 +\frac{1}{4b}\int dx\str (\nabla Q(x))^2
 -\frac{1}{4c}
 \int dx\left(\str Q(x)\nabla Q(x)\Sigma_x\right)^2, \label{F1}
\ee
 where for the sake of simplicity we have neglected the source term, and 
 $V=n$ is the 'system volume'. 
 The coupling constants $b$ and $c$ are given by
\be
 \frac{1}{b} = -\frac{A_0^{(2)}}{A_0}
 = r_0^2\frac{\sin\frac{\pi}{2\alpha}}{\sin\frac{3\pi}{2\alpha}}, 
 \quad
 \frac{1}{c} = -\frac{A_0^{(2)}}{4A_0^2}
 = \frac{r_0}{8}
 \frac{\sin^2\frac{\pi}{2\alpha}}{\frac{\pi}{2\alpha}\sin\frac{3\pi}{2\alpha}}.
\ee
 For $1/2 < \alpha \leq 3/2$
 the gradient expansion requires a special care since 
 $A_0^{(2)}$ is not well-defined.
 Taking into account higher order terms in the gradient expansion 
 (as in Ref.\cite{prbm}),
 we find the following expression for $A_0^{(2)}$,  
\be
 q^2\int_{0}^\infty dr\frac{r^2}{1+(r/r_0)^{2\alpha}} 
 \to 2\int_{0}^\infty dr
 \frac{1-\cos qr}{1+(r/r_0)^{2\alpha}} 
 \sim C_\alpha r_0^{2\alpha-1}|q|^{2\alpha-1},
\ee
 where $C_\alpha=2\int_0^\infty dx (1-\cos x)/x^{2\alpha}$
 is a numerical constant.
 With this substitution 
 Eq.(\ref{F0}) for $1/2 < \alpha\leq 3/2$ is given by 
\be
 F &=&  \frac{i\pi\e}{2\D V}\int dx\str \Sigma_zQ(x) 
 +\frac{1}{4b}\int \frac{dq}{2\pi} |q|^{2\alpha-1}\str Q(q)Q(-q) \no\\
 & & -\frac{1}{4c}
 \int\frac{dq}{2\pi}\frac{dp}{2\pi}\frac{dp'}{2\pi}|q|^{2\alpha-1}
 \str Q(p+q)Q(-p)\Sigma_x \str Q(p'-q)Q(-p')\Sigma_x, \label{F2}
\ee
 where 
\be
 \frac{1}{b} = \frac{2C_\alpha r_0^{2\alpha}}{A_0}
 = C_\alpha r_0^{2\alpha-1}
 \frac{\sin\frac{\pi}{2\alpha}}{\frac{\pi}{2\alpha}}, 
 \quad
 \frac{1}{c} = \frac{C_\alpha r_0^{2\alpha}}{2A_0^2}
 = \frac{C_\alpha r_0^{2\alpha-2}}{8}
 \left(\frac{\sin\frac{\pi}{2\alpha}}{\frac{\pi}{2\alpha}}\right)^2.
\ee
 In this case, unlike for $\alpha>3/2$,
 the nonlinear $\sigma$ mode (\ref{F2}) corresponds with 
 a process of anomalous diffusion.
 However, we note that 
 the relation $1/b=4A_0/c\sim r_0/c$ holds in both cases
 and  $1/c \ll 1/b$. 
 Thus the double trace term (last term of $F$)
 will not affect the spectral properties to leading order in $1/r_0$.
 Its role in the localization properties of the chRBM is discussed below.

\subsection{Renormalization group equations}

 By using the above nonlinear $\sigma$ model 
 Eqs.(\ref{F1}) and (\ref{F2}),
 one can discuss the localization properties of our chRBM 
 in the thermodynamic limit. 
 This can be done by investigating the running of 
 the coupling constants $b$ and $c$ under the RG flow.
 We compute the RG equations to leading order 
 in $b$ and $b^2/c$.
 For the normal diffusive case, Eq.(\ref{F1}), 
 the parameters $b$ and $b^2/c$ involve the following propagators  
\be
 \Pi(q,\omega) = \frac{1}{q^2-i\omega}, \quad
 \Pi_2(q,\omega) = q^2\Pi^2(q,\omega) \quad (\mbox{For\ } \alpha> 3/2),
\ee
 respectively.
 On the other hand, in the anomalous diffusive case, Eq.(\ref{F2}), 
 the propagators are given by 
\be
 \Pi(q,\omega) = \frac{1}{|q|^{2\alpha-1}-i\omega}, \quad
 \Pi_2(q,\omega) = |q|^{2\alpha-1}\Pi^2(q,\omega)
 \quad (\mbox{For\ } 1/2<\alpha\leq 3/2).
\ee
 By observing the momentum integrations of the propagators 
\be
 \int_{\Lambda/l}^\Lambda \frac{d^dq}{(2\pi)^d}\Pi(q,\omega) 
 \sim \frac{1}{2\pi}\ln l,\quad
 \int_{\Lambda/l}^\Lambda \frac{d^dq}{(2\pi)^d} \Pi_2(q,\omega) 
 \sim \frac{1}{2\pi}\ln l,
\ee
 where $\Lambda$ is a cutoff and $l\sim n>1$, 
 we conclude that for normal diffusion the logarithmic dimension is $d=2$.
 For anomalous diffusion, since our chRBM is in essence one dimensional,
 the logarithmic dimension corresponds to the case $\alpha =1$ 
 (generally, $d=2\alpha-1$).
 We note that it is a straightforward task to extend 
 the derived $\sigma$ model for the chRBM to arbitrary dimension.

 In the normal diffusive case, 
 the perturbative calculation of the RG equations was 
 first done in Ref.\cite{gade} by the method of the $\e$-expansion.
 The RG equations are given by 
\be
 \beta_b = -\frac{db}{d\mu} = \e b, \quad
 \beta_c = -\frac{dc}{d\mu} = \e c+\frac{1}{8\pi}c^2, \quad
 \zeta = \frac{1}{2\pi}\frac{b^2}{c},
\ee
 where $\mu=\ln l$ is the renormalization scale,
 $\beta_{b,c}$ are the beta functions for $b$ and $c$ and  
 $\zeta$ is the wavefunction renormalization.
 The above equations are immediately applied to 
 the anomalous case by replacing $\e=d-2$ by $\e=1-(2\alpha-1)$.
 The RG equations are easily solved for arbitrary $\e$ \cite{gade}.
 We see that, in the domain $\e>0$, the system is within the extended phase.
 For $\e<0$, $b$ and $b^2/c$ goes to infinity and 
 consequently a transition to localization is expected.
 However we note that in this limit 
 the perturbative expansion breaks down 
 since $b,c$ increase under the RG flow.
 For $\e=0$, 
 $b$ remains constant, $c$ goes slowly ($\sim 1/\log l$) to zero, 
 and $b^2/c$ goes to infinity.
 That means that at the origin and for exponentially large volumes
 the eigenstates should be delocalized and 
 the DoS is expected to diverge. 
 Of course this divergence is somewhere cut off before the origin 
 since random matrix theory 
 (which should be valid in the deep infrared limit) 
 predicts a vanishing of the DoS at the origin.

 Finally we remark that the above pseudo-divergence is 
 beyond the current numerical capabilities since  $b\ll c$ 
 and  $1/c \sim \log l \sim \log n$ for $n\gg 1$. 

\section{Level correlation functions}
\label{level}

 In this section we compute the DoS and the TLCF close to the origin 
 in the $g \gg 1$ limit by using the nonlinear $\sigma$ model 
 introduced in the previous section.
 As mentioned previously, in systems with chiral symmetry 
 there is a clear distinction between the spectral correlations
 near zero (the origin) and away from zero (the bulk).
 In the bulk the spectral correlations are not affected by the block 
 structure and coincide with the nonchiral version of Eq.(\ref{rb}) 
 which has been intensively studied in recent years \cite{prbm,prbm2}.
 After reviewing the result in the bulk, 
 we examine the origin for the anomalous diffusion case $1/2<\alpha\leq 3/2$.  
 For normal diffusion, we refer to Ref.\cite{ktaka}.
 The case $\alpha=1$, related to the Anderson transition, 
 is worked out in detail.

\subsection{Review of results at the bulk}

 The use of the supersymmetry method permits 
 an analytical evaluation of both spectral properties and 
 eigenfunction statistics \cite{prbm} in a certain region of parameters. 
 In the thermodynamic limit it was found that the eigenfunctions are 
 multifractal for $\alpha =1$ and 
 localized (delocalized) for $\alpha >1$ ($\alpha <1$) \cite{prbm}.
 The spectral correlations in the limit $g=\pi E_c/\Delta \gg 1$ 
 ($E_c$ the Thouless energy)
 can be expressed through the spectral determinant of 
 a classical diffusion operator \cite{AA}.
 For the unitary ensemble 
\be
 R(z) &=& 1-\frac{1}{2z^2}
 -\frac{1}{4}\frac{\partial^2}{\partial z^2} \ln {\cal D}^2(z)
 +\frac{\cos(2z)}{2z^2} {\cal D}^2(z),  \label{aa}\\
 {\cal D}(z) &=& \prod_{n=1}^\infty
 \frac{g^2(2\pi n)^{2(2\alpha-1)}}{g^2(2\pi n)^{2(2\alpha-1)}+z^2},
 \label{sdet}
\ee
 where $R(z)=\Delta^2\langle\rho(\epsilon_1)\rho(\epsilon_2)\rangle-1$ 
 is the TLCF, $\rho(\epsilon)$ is the DoS at energy $\epsilon$, 
 $\Delta$ is the mean level spacing, $1/2<\alpha\le 3/2$, and 
 the energy $z$ is expressed in units of $\Delta$ as $z=\pi(\e_1-\e_2)/\D$.
 Although this result was derived assuming $z\gg 1$, 
 in Ref.\cite{MI} it was shown that it is valid for all $z$. 

 We mention that the spectral determinant above 
 corresponds with a process of anomalous diffusion \cite{MK}  
 with $\langle|r|\rangle\propto t^{1/(2\alpha-1)}$. 
 The asymptotic behavior of $R(z)\sim z^{-2+1/(2\alpha-1)}$ 
 ($\alpha \neq 1$) is power-law and the dimensionless conductance
 $g \sim n^{2-2\alpha}$ increases (decreases) 
 with the system size $n$ for $\alpha<1$ ($\alpha>1$).
 Both the scaling of $g$ and the TLCF resemble those of a weakly disordered 
 conductor in $d=2/(2\alpha -1)$ dimensions \cite{prbm}.
 For the special case $\alpha=1$,
 ${\cal D}(z)=(z/2g)\sinh^{-1}(z/2g)$ 
 and $R(z)$ can be explicitly evaluated as
\be
 R(z) = 1-\frac{\sin^2 z}{z^2}
 \left(\frac{z/2g}{\sinh (z/2g)}\right)^2. \label{aa1}
\ee
 This correlation function reproduces typical features 
 of the spectral correlations at the Anderson transition as 
 level repulsion for $z \ll 1$ 
 but sub-Poisson number variance $\Sigma^2(L) \sim \chi L$ for $L \gg 1$ 
 (see Sec.\ref{comp} for a definition of the number variance).

 Although the RBM is defined in 1D, 
 generalizations to $d$ dimensions are straightforward. 
 In that case the product over $n$ runs over all possible 
 ${\vec n}=(n_1,\ldots,n_d)$.
 As mentioned previously, according to Levitov's results \cite{levitov}, 
 the properties of these power-law hopping models are similar 
 in different dimensions provided that the dimension of the space 
 is equal to the power-law decay. 
 It is tempting to guess that a power-law decay $\alpha$ in $d=1$ 
 is similar to a decay $d\alpha$ in $d$ dimensions.

\subsection{Results at the origin}

 After reviewing the properties at the bulk 
 we move to the spectral correlations close to the origin.
 In the limit 
 $g=\pi E_c/\D=n^{2-2\alpha}/b\sim r_0^{2\alpha-1}n^{2-2\alpha}\gg 1$, 
 we calculate the DoS and the TLCF of the chRBM (\ref{rb2})
 by using the nonlinear $\sigma$ model as given by Eq.(\ref{F2}) 
 without the double trace term.
 As mentioned previously, since $b/c \sim 1/r_0 \ll 1$, 
 the double-trace term can be safely neglected.

 The DoS is given by the expression
\be
 \left<\rho(\e)\right> 
 = \frac{1}{4\Delta V}\mbox{Re}\int {\cal D}Q
 \left[\int dx\str k\Si_z Q(x)\right] \mbox{e}^{-F},
\ee
 where $F$ is defined in Eq.(\ref{F2}).
 The perturbative calculation corresponds to expand 
 the $Q$-matrix as 
\be
  Q(x) = \Si_z\frac{1+iP}{1-iP}
  = \Si_z\left(1+2iP-2P^2+\cdots\right), \quad
  P = \left(\ba{cc} 0 & t \\ t & 0 \ea\right), \quad
  t = \left(\ba{cc} a & \sigma \\ \rho & ib \ea\right), 
\ee
 where $a$,$b$ are real variables and $\sigma$,$\rho$ Grassmann variables.
 The resulting expansion is in powers of 
 the (anomalous) diffusion propagator
\be
 \label{pa}
 \Pi(q,\e) = \frac{\D}{2\pi}\frac{1}{D|q|^{2\alpha-1}-i\e},
\ee
 where $D=\D n/\pi b$.
 Finite $g$ corrections to the DoS are given by
\be
 \left<\rho(\e)\right> 
 \sim \frac{1}{\D}
 \biggl[1+\frac{1}{2}\mbox{Re}\Bigl(\sum_{q}\Pi(q,\e)\Bigr)^2
 \biggr].
\ee

 Since the expansion is not well-defined for the zero mode ($q=0$), 
 this contribution should be removed from the above expression and 
 treated in a separate way \cite{KM}.
 We remark, as was noticed in Ref.\cite{ktaka}, 
 the mean level spacing $\D$ corresponding to 
 the ergodic regime is modified by a finite $g$ as
\be
 \frac{1}{\tilde{\Delta}} &=& 
 \frac{1}{\Delta}\mbox{Re}\int {\cal D}\tilde{Q}
 \left[\frac{1}{4V}\int dx\str k\Si_z \tilde{Q}(x)\right]
 \mbox{e}^{-F[\tilde{Q}]}
 \sim \frac{1}{\D}
 \biggl[1+\frac{1}{2}\mbox{Re}\Bigl(\sum_{q\ne 0}\Pi(q,\e)\Bigr)^2
 \biggr], \label{tildedelta}
\ee
 where $\tilde{Q}$ denotes the nonzero modes.
 This quantity is relevant for us 
 since we are interested in level correlations of unfolded variables.
 Thus we define the DoS scaled in terms of 
 the renormalized mean level spacing
\be
 \rho_1(z) = \tilde{\Delta}\langle\rho(\e=\tilde{\D}z/\pi)\rangle. \label{rho1}
\ee
 We note that, as shown in Sec.\ref{comp}, 
 the renormalization of $\Delta$ is important 
 to find agreement between the results of this section 
 and those of Ref.\cite{ant1}.

 The zero mode contribution is treated nonperturbatively 
 and gives the ergodic result $g\to\infty$ \cite{VZ} 
\be
 \rho_1(z) \to \rho_1^{(0)}(z) 
 = \frac{\pi z}{2}(J_0^2(z)+J_1^2(z)). \label{msd}
\ee
 As in the nonchiral case \cite{efetov}, 
 a proper parametrization of the $Q$-matrix 
 is required to derive this expression.
 It can be incorporated into our $Q$-matrix as
\be
 Q(x) = T\tilde{Q}(x)\bar{T},
\ee
 where the supermatrix $T$ ($\bar{T}$ is the inverse of $T$) parametrizes 
 the zero mode as \cite{AST}
\be
 T &=& UT_0\bar{U}, \no\\
 T_0 &=& 
 \left(\ba{cc} \cos\frac{\hat{\theta}}{2} & -i\sin\frac{\hat{\theta}}{2} \\
 -i\sin\frac{\hat{\theta}}{2} & \cos\frac{\hat{\theta}}{2} \ea\right), \quad
 \hat{\theta} = \left(\ba{cc} \theta_F & 0 \\ 0 & i\theta_B \ea\right), \no\\
 U &=& \left(\ba{cc} u & 0 \\ 0 & u \ea\right), \quad
 u = \exp\left(\ba{cc} 0 & \xi \\ \eta & 0 \ea\right), \label{nonpert}
\ee
 where $-\pi\le\theta_F\le\pi$, $0\le\theta_B\le\infty$,
 and $\xi$ and $\eta$ are Grassmann variables.
 Using this parametrization, 
 we integrate the zero mode first and, then,
 treat the nonzero modes perturbatively.
 This was done in Ref.\cite{ktaka} for the case of normal diffusion.
 In the present case, the zero mode part is unchanged 
 since the kinetic term, second (and third) term 
 in the free energy (\ref{F2}), does not include the zero mode.
 For the nonzero mode, the anomalous diffusion propagator Eq.(\ref{pa})
 is used for the perturbative expansion.
 Thus after integration of the zero mode, we are left with,
\be
 \label{d1}
 \left<\rho(\e)\right> &\sim& 
 \frac{1}{\tilde{\Delta}}
 +\frac{\pi}{2\Delta}\mbox{Re}\frac{d}{dz_0}
 \int_{z_0}^\infty dt(t-z_0)
 \left<J_0(t+z_0A_B)J_0(t-z_0A_F)
 -J_1(t+z_0A_B)J_1(t-z_0A_F)\right>_{\rm kin}, \no\\
 A_{F,B} &=& 
 -\frac{1}{2V}\int dx\str \frac{1\pm k}{2}\Si_z [\tilde{Q}(x)-\Si_z], \no\\
 \left<\cdots\right>_{\rm kin} &=& \int {\cal D}\tilde{Q}
 \left(\cdots\right)
 \exp\left[-\frac{1}{4b}\int\frac{dq}{2\pi} |q|^{2\alpha-1}
 \str \tilde{Q}(q)\tilde{Q}(-q)\right],
\ee
 where $z_0=\pi\e/\Delta$ and 
 $1/\tilde{\Delta}$ is the purely perturbative contribution (\ref{tildedelta}). 
 We note that this expression is valid up to second order in $1/g$.
 The ergodic result (\ref{msd}) can be easily recovered by putting $A_{F,B}=0$.

 We are now ready to evaluate Eq.(\ref{d1}) in different domains. 
 In all cases the limit $g \gg 1$ is assumed.
 For $z\ll g$, known as the Kravtsov-Mirlin (KM) domain \cite{KM},
 the Bessel functions can be expanded
 in powers of $zA\sim z/g$ as
\be
 J(t-zA) \sim \left(1-zA\frac{d}{dt}
 +\frac{z^2A^2}{2}\frac{d^2}{dt^2}\right)J(t),
\ee
 to obtain 
\be
 \left<\rho(\e)\right> &\sim& 
 \frac{1}{\Delta}\mbox{Re}
 \biggl[1+\frac{1}{2}\left<A_B-A_F\right>_{\rm kin}\frac{d}{dz_0}z_0 
 +\frac{1}{8}\left<(A_B-A_F)^2\right>_{\rm kin}
 \frac{d}{dz_0}z_0^2\frac{d}{dz_0}\biggr]
 \rho_1^{(0)}(z_0). \no\\
\ee
 Changing the variable from $z_0=\pi\e/\D$ to $z=\pi\e/\tilde{\D}$,
 we finally are left with
\be
 \rho_1(z) \sim
 \left[1+\frac{a_\alpha}{8g^2}
 \left(2z\frac{d}{dz}+z^2\frac{d^2}{dz^2}\right)
 \right]\rho_1^{(0)}(z), \label{doskm}
\ee
 where $\rho_1^{(0)}(z)$ is defined in Eq.(\ref{msd}) and
 $a_\alpha$ is the momentum summation for periodic boundary conditions,
\be
 a_\alpha = g^2\sum_{q\ne 0}\Pi^2(q,0) 
 = 2\sum_{n=1}^\infty\frac{1}{(2\pi n)^{4\alpha-2}}.
\ee

 We now move to the Andreev-Altshuler (AA) domain \cite{AA} $z\gg 1$.
 In this limit we cannot expand the Bessel functions.
 Instead we use their asymptotic form,
\be
 J_0(z) = \frac{1}{\pi}\int_0^\pi dx\mbox{e}^{iz\cos x} 
 \sim \sqrt{\frac{1}{\pi z}}\biggl[
 \left(1-\frac{1}{8z}+\cdots\right)\cos z 
 +\left(1+\frac{1}{8z}+\cdots\right)\sin z\biggr].
\ee
 Noting that $J_1(z)=-J_0'(z)$ we find
\be
 \rho_1(z) &\sim& 
 \mbox{Re}\left[
 1+\frac{1}{8z^2}
 -\frac{1}{2z}
 \mbox{e}^{2iz}\int{\cal D}\tilde{Q}
 \mbox{e}^{-F_k(z)}\right], \\
 F_k(z) &=& 
 \frac{1}{4b}\int\frac{dq}{2\pi} |q|^{2\alpha-1}
 \str \tilde{Q}(q)\tilde{Q}(-q)
 -\frac{iz}{2V}\int dx\str k\Si_z (\tilde{Q}-\Si_z).
\ee
 We remark that the presence of the supermatrix $k=\mbox{diag}(1,-1)$ 
 breaks the supersymmetry in $F_k(z)$.
 Keeping terms up to second order in the $P$-matrix
 we obtain
\be
 \rho_1(z) \sim
 1-\frac{\cos 2z}{2z}{\cal D}(z)+\frac{1}{8z^2}, \label{dosaa}
\ee
 where the spectral determinant ${\cal D}(z)$ is given 
 by Eq.(\ref{sdet}).
 We note that the ergodic limit corresponds to put ${\cal D}=1$
 which corresponds with the asymptotic form of the exact result Eq.(\ref{msd}).

 We now consider the TLCF. 
 In this case the analytical calculation 
 is along the lines of the DoS though more technically involved. 
 Here we present a summary of results. 
 For technical details we refer to Ref.\cite{ktaka} 
 where the case of normal diffusion was discussed.
 We start by defining 
\be
 \rho_2(z_1,z_2) =
 \tilde{\D}^2\langle\langle
 \rho(\e_1=\tilde{\D} z_1/\pi)\rho(\e_2=\tilde{\D} z_2/\pi)
 \rangle\rangle,
 \label{rho2}
\ee
 where $\langle\langle\rho(\e_1)\rho(\e_2)\rangle\rangle
 =\langle\rho(\e_1)\rho(\e_2)\rangle
 -\langle\rho(\e_1)\rangle\langle\rho(\e_2)\rangle$ 
 is the connected part of the TLCF.
 In the KM's domain $z_{1,2}\ll g$, 
\be
 \rho_2(z_1,z_2) 
 &\sim& -\left\{\left[
 1+\frac{a_\alpha}{4g^2}
 \left(z_1\frac{\partial}{\partial z_1}
 +z_2\frac{\partial}{\partial z_2}\right) 
 +\frac{a_\alpha}{8g^2}\left(z_1^2\frac{\partial^2}{\partial z_1^2}
 +2z_1z_2\frac{\partial}{\partial z_1}\frac{\partial}{\partial z_2}
 +z_2^2\frac{\partial^2}{\partial z_2^2}\right)
 \right]K(z_1,z_2)\right\}^2, \label{2ptkm}
\ee
 where 
\be
 K(z_1,z_2) = \frac{\pi\sqrt{z_1z_2}}{z_1^2-z_2^2}
 (z_1J_1(z_1)J_0(z_2)-z_2J_0(z_1)J_1(z_2)). \label{K}
\ee
 The ergodic limit is given by $\rho_2(z_1,z_2)=-K^2(z_1,z_2)$.
 In the AA's domain $z_{1,2}\gg 1$, 
\be
 \rho_2(z_1,z_2) &\sim& 
 \frac{1}{2}\mbox{Re}\sum_{q^2\ne 0}\left(\Pi_+^2+\Pi_-^2\right) 
 +\frac{\sin 2z_1}{2z_1}{\cal D}_1 \mbox{Im}\sum_{q^2\ne 0}
 \left(\Pi_++\Pi_-\right) 
 +\frac{\sin 2z_2}{2z_2}{\cal D}_2 \mbox{Im}\sum_{q^2\ne 0}
 \left(\Pi_+-\Pi_-\right) \no\\
 & & +\frac{1}{8z_1z_2}\left[
 {\cal D}_1{\cal D}_2({\cal D}_+^{2}{\cal D}_-^{-2}-1)
 \cos 2(z_1+z_2) 
 +{\cal D}_1{\cal D}_2({\cal D}_-^{2}{\cal D}_+^{-2}-1)
 \cos 2(z_1-z_2)
 \right] \no\\
 & & -\frac{1}{2(z_1+z_2)^2}\left[1+
 {\cal D}_1{\cal D}_2{\cal D}_+^2{\cal D}_-^{-2}\cos 2(z_1+z_2)\right]
 -\frac{1}{2(z_1-z_2)^2}\left[1-
 {\cal D}_1{\cal D}_2{\cal D}_-^2{\cal D}_+^{-2}\cos 2(z_1-z_2)\right]  \no\\
 & & +\frac{1}{z_1^2-z_2^2}\left(
 {\cal D}_1  \sin 2z_1-{\cal D}_2 \sin 2z_2\right), 
 \label{2ptaa}
\ee
 where ${\cal D}_{1,2}={\cal D}(z_{1,2})$, 
 ${\cal D}_{\pm}={\cal D}((z_1\pm z_2)/2)$, Eq.(\ref{sdet}), and 
 $\Pi_{1,2}=\Pi(q,\e_{1,2})$, $\Pi_\pm=\Pi(q,(\e_1\pm \e_2)/2)$, Eq.(\ref{pa}).
 We remark that the first term of the r.h.s. is obtained by 
 a purely perturbative calculation.

 We stress that in the the common domain $1\ll z \ll g$ 
 the results in both regions AA and KM coincide.
 It is worthwhile to note that in the unitary limit $z,z_1+z_2\to\infty$
 our results also coincide with those of the nonchiral version of our model:
 $\rho_1(z)\to 1$ and 
\be
 R(z_1,z_2) = 1+\frac{\rho_2(z_1,z_2)}
 {\rho_1(z_1)\rho_1(z_2)}
 \to 1+\frac{1}{2}\mbox{Re}\sum_{q^2\ne 0}\Pi_-^2
 -\frac{1}{2(z_1-z_2)^2}
 +\frac{\cos 2(z_1-z_2)}{2(z_1-z_2)^2}{\cal D}_-^2.
\ee
 Noting $\Pi(q,\e/2;g)=2\Pi(q,\e;2g)$ and ${\cal D}(z/2;g)={\cal D}(z;2g)$, 
 we see that this is the AA's result (\ref{aa})
 for the unitary class where $g$ is substituted by $2g$.
 The factor 2 is considered to be due to the chiral symmetry.

\subsection{Explicit results for $\alpha=1$}
\label{alpha1}

 The case $\alpha=1$ is specially interesting. 
 On the one hand it is related to the spectral correlations 
 at the Anderson transition and on the other hand 
 the analytical results of the previous section are greatly simplified 
 since the spectral determinant can be evaluated exactly.

 In the KM domain the summation in $a_\alpha$ can be easily performed,
\be
 a_1 = \frac{1}{2\pi^2}\sum_{n=1}^\infty\frac{1}{n^2}=\frac{1}{12}.
\ee
 The DoS and the TLCF are consequently given by
\be
 \rho_1(z) &\sim&
 \left[1+\frac{1}{96g^2}
 \left(2z\frac{d}{dz}+z^2\frac{d^2}{dz^2}\right)
 \right]\rho_1^{(0)}(z), \\
 \rho_2(z_1,z_2) 
 &\sim& -\left\{\left[
 1+\frac{1}{48g^2}
 \left(z_1\frac{\partial}{\partial z_1}
 +z_2\frac{\partial}{\partial z_2}\right) 
 +\frac{1}{96g^2}\left(z_1^2\frac{\partial^2}{\partial z_1^2}
 +2z_1z_2\frac{\partial}{\partial z_1}\frac{\partial}{\partial z_2}
 +z_2^2\frac{\partial^2}{\partial z_2^2}\right)
 \right]K(z_1,z_2)\right\}^2.
\ee
 The results in AA's domain $1\ll z_{1,2}$, $1\ll g$ are greatly 
 simplified since both the spectral determinant 
\be
\label{sdet1}
 {\cal D}(z) = \prod_{n=1}^\infty 
 \frac{1}{1+(z/2\pi gn)^2}
 = \frac{z/2g}{\sinh\left(z/2g\right)},
\ee
 and the momentum summations of the propagator can be exactly evaluated,
\be
 \label{pa1}
 & & \mbox{Re}\sum_{q^2\ne 0}\Pi^2(q,\e) =
 \frac{1}{8\pi^2 g^2}\mbox{Re}\sum_{n=1}^\infty
 \frac{1}{\left(n-iz/2\pi g\right)^2} 
 =  \frac{1}{4z^2}\left(1-{\cal D}^2(z)\right), \no\\
 & & \mbox{Im}\sum_{q^2\ne 0}\Pi(q,\e) = 
 \frac{i}{4\pi g}\sum_{n=1}^\infty
 \left(\frac{1}{n+iz/2\pi g}-\frac{1}{n-iz/2\pi g}\right)
 = -\frac{1}{2z}\left(1-\frac{z}{2g}\coth\frac{z}{2g} \right).
\ee
 By utilizing the relations  
\be
 \frac{1}{2g}\left(
 \coth \frac{z_1+z_2}{4g}+\coth \frac{z_1-z_2}{4g}
 \right) &=& 
 \frac{4z_1}{z_1^2-z_2^2}{\cal D}_1^{-1}{\cal D}_+{\cal D}_-,
 \no\\
 {\cal D}_1{\cal D}_2({\cal D}_+^{2}{\cal D}_-^{-2}-1)
 &=& -\frac{4z_1z_2}{(z_1-z_2)^2}({\cal D}_+^2-{\cal D}_1{\cal D}_2), 
\ee
 we find the following simple form for the TLCF,
\be
 \rho_2(z_1,z_2)
 = -\left[\frac{\sin (z_1-z_2)}{z_1-z_2}{\cal D}_-
 -\frac{\cos (z_1+z_2)}{z_1+z_2}{\cal D}_+\right]^2. \label{2ptaa1}
\ee 
 We easily see that the unitary limit 
 $z_1+z_2\to\infty$, Eq.(\ref{aa1}),
 can be recovered by keeping only the first term in the bracket.

\section{Chiral symmetry and Anderson transition}
\label{comp}

 In this section, for the special value $\alpha=1$, 
 we put forward an exact result for the DoS and the TLCF, 
 valid for all $z,z_{1,2}$, in the limit $g \gg 1$.
 The spectral fluctuations in this case present features as scale invariance,
 level repulsion, and asymptotically linear number variance similar
 to the ones at the Anderson transition. 
 Thus the chRBM at $\alpha=1$ is an ideal candidate to investigate  
 the interplay between chiral symmetry and wavefunction localization 
 leading to the Anderson transition.
 We recall that in the thermodynamic limit, 
 according to the RG analysis of previous sections,
 the chiral symmetry will delocalize the eigenstates close to the origin
 and the DoS is divergent (though according to chiral random matrix 
 prediction this divergence is cut off before the origin).

\subsection{Generalized chiral random matrix model for critical statistics}
 
 We find exact expressions for the DoS and TLCF
 of the chRBM at $\alpha=1$
 by mapping it onto an exactly solvable generalized chiral
 random matrix model (gchRMM).  
 The model in question is defined by the probability density 
\be
 \label{qw}
 P({\cal H}) \propto  
 \int {\cal D}U \exp\tr\left[
 -\frac{1}{4h}{\cal H}^2 
 -n^2h[{\cal H},U] [{\cal H},U]^\dag
 \right], 
\ee
 where $\cal H$ is given by Eq.(\ref{rb}), 
 $U$ has the chiral block structure
\be
 U = \left(\ba{cc} U_1 & 0 \\ 0 & U_2 \ea\right),
\ee
 with $U_{1,2}$ $n\times n$ unitary matrices, and 
 ${\cal D}U$ denotes the Haar measure of the unitary matrices $U_{1,2}$.
 We shall see that the parameter $h$ 
 is indeed related to the conductance $g$ by $h\sim 1/g$.
 In Refs.\cite{KT,ant3}, it was found that the spectral correlations of 
 this model are equivalent to the spatial correlations of 
 the Calogero-Sutherland (CS) model \cite{CS} at finite temperature. 
 By using this analogy, exact expressions for 
 the spectral correlations were found for any $g \sim 1/h$.
 It turns out that in the $g \gg 1$ limit 
 such model is also equivalent to our chRBM at $\alpha=1$ \cite{mirlin}. 
 The argument is as follows.
 By decomposing Eq.(\ref{qw}) into the blocks of ${\cal H}$ and $U$, 
 the probability density can be written as
\be
 P({\cal H}) \propto  
 \int {\cal D}U \exp\tr\left[
 -\frac{1+8h^2n^2}{2h}CC^\dag
 +4n^2h {\rm Re}U_1 C U_2 C^\dag \right].
 \label{ZUCVC}
\ee
 The integral over $U$ is, in principle, performed by using 
 the Harish-Chandra-Itzykson-Zuber formula. 
 However one could attempt an alternate route. 
 Random matrix theory predicts that 
 the spectral density of $U$ in the large $n$ limit is constant and, 
 due to level repulsion, the eigenvalues are consequently well separated. 
 One can thus model the spectrum typical of a $U$ as $\exp(i{\theta_{m}})$ 
 ($m=1, \ldots 2n$) where in a first approximation 
 the phases $\theta_m=2\pi m/2n$ are assumed to be equidistant.
 By doing that we are neglecting the small fluctuations 
 around the equidistant position.
 Within these approximations 
 the integral over $U$ is reduced, up to multiplicative
 constant factors, to the contribution of one ``typical'' $U$.
 In a basis such that the $U_{1,2}$ in (\ref{ZUCVC}) becomes diagonal,
\be
 P({\cal H}) \propto 
 \exp\left\{
 \sum_{i,j}|C_{ij}|^2\left[1+(4nh)^2\sin^2{\frac{\pi(i-j)}{2n}}\right]
 \right\}.
\ee
 This expression describes a chRBM with variance 
\be
 a^2(r=|i-j|)\sim 
 \frac{1}{1+(4nh)^2\sin^2\frac{\pi r}{2n}}
 \sim
 \frac{1}{1+4\pi^2h^2r^2},
\ee 
 but this is nothing but our original chRBM for $\alpha =1$
 with $r_0 \sim 1/h$.
 Thus the spectral correlations of both models should coincide. 
 Let us first review the exact results 
 for the gchRMM reported in Ref.\cite{ant1}.
 In the $h \ll 1$ limit, 
 the DoS was found to be
\be
 \label{dos1}
 \langle\rho(\e)\rangle = 
 \frac{1}{\D}\int_{-1/h}^\infty dt
 \frac{\left(1+ht\right)^2}{\cosh^2 \left[2t(1+\frac{ht}{2})\right]}
 \rho_1^{(0)}\left((1+ht)\frac{\pi\e}{\D}\right), 
\ee
 where $\Delta=\pi/2n$ is the mean level spacing at $h=0$, and 
 $\rho_1^{(0)}(z)$ is given by Eq.(\ref{msd}).
 The TLCF is 
\be
\label{tlcf1}
 \langle\langle\rho(\e_1)\rho(\e_2)\rangle\rangle 
 = -\frac{1}{\D^2}
 \biggl[
 \int_{-1/h}^\infty dt
 \frac{\left(1+ht\right)^2}{\cosh^2 \left[2t(1+\frac{ht}{2})\right]}
 K\left((1+ht)\frac{\pi\e_1}{\D},(1+ht)\frac{\pi\e_2}{\D}\right)
 \biggr]^2,
\ee
 where $K(z_1,z_2)$ is given by Eq.(\ref{K}).

 The mean level spacing $\D(h)$ at $h\ne 0$ 
 can be evaluated from Eq.(\ref{dos1}) as 
\be
 \frac{1}{\D(h)} = \frac{1}{\D}
 \int_{-1/h}^\infty dt
 \frac{\left(1+ht\right)^2}{\cosh^2 \left[2t(1+\frac{ht}{2})\right]}
 \sim  \frac{1}{\D}\left(1-\frac{\pi^2h^2}{96}\right). \label{deltah}
\ee
 We note that this quantity corresponds to $\tilde{\D}$
 in Eq.(\ref{tildedelta}).
 Naive calculation of $\tilde{\D}$ using the nonlinear $\sigma$ model
 gives a divergent result and 
 we thus find disagreement with Eq.(\ref{deltah}). 
 However, just like in Eqs.(\ref{rho1}) and (\ref{rho2}),
 if the above results are scaled in terms of $\D(h)$ as
\be
 \rho_1(z) &=& \D(h)\langle\rho(\e=\D(h)z/\pi)\rangle, \no\\
 \rho_2(z_1,z_2) &=& 
 \D^2(h)\langle\langle
 \rho(\e_1=\D(h) z_1/\pi)\rho(\e_2=\D(h) z_2/\pi)
 \rangle\rangle,
\ee
 we show below that agreement between both calculations is found.

 In order to prove this claim
 we perform a series expansion on 
 the exact results (\ref{dos1}) and (\ref{tlcf1})
 to compare them with the findings of the previous section 
 in both the AA and KM domain. 
 For the KM domain we start with the expression (\ref{dos1}).
 The integral is strongly peeked around $t=0$. 
 Therefore we can perform an expansion in powers of $t$ 
 up to terms involving $h^2$ corrections.
\be
 \langle\rho(\e)\rangle &\sim& 
 \frac{1}{\D}\int_{-1/h}^\infty dt
 \frac{1}{\cosh^2 2t}\left(
 1+2ht-2ht^2\frac{\sinh 2t}{\cosh 2t}
 +h^2t^2
 -4h^2t^3\frac{\sinh 2t}{\cosh 2t}
 -h^2t^4
 +3h^2t^4\frac{\sinh^2 2t}{\cosh^2 2t}
 \right) \no\\
 & & \times
 \left(1+htz\frac{d}{dz}
 +\frac{h^2t^2}{2}z^2\frac{d^2}{dz^2}
 +\frac{\pi^2 h^2}{96}z\frac{d}{dz}\right)\rho_1^{(0)}(z) \no\\
 &\sim& 
 \frac{1}{\D(h)}\left[
 1+\frac{\pi^2 h^2}{96}
 \left(2z\frac{d}{dz}+z^2\frac{d^2}{dz^2}\right)
 \right]\rho_1^{(0)}(z).
\ee
 The TLCF is calculated in the same way.
 After a laborious calculation one is left with 
\be
 \langle\langle\rho(\e_1)\rho(\e_2)\rangle\rangle 
 &\sim& 
 -\frac{1}{\D^2(h)}
 \biggl\{\biggl[
 1+\frac{\pi^2 h^2}{48}\left(z_1\frac{d}{dz_1}+z_2\frac{d}{dz_2}\right) \no\\
 & & 
 +\frac{\pi^2 h^2}{96}\left(z_1^2\frac{d^2}{dz_1^2}
 +2z_1z_2\frac{d}{dz_1}\frac{d}{dz_2}
 +z_2^2\frac{d^2}{dz_2^2}\right)
 \biggr]K(z_1,z_2)\biggr\}^2.
\ee
 In the AA's domain $1\ll z,z_{1,2}$,
 the DoS reduces to 
\be
 \langle\rho(\e)\rangle &\sim& \frac{1}{\D(h)}\int_{-\infty}^\infty dt
 \frac{1}{\cosh^2 2t}
 \left[1-\frac{\cos \left[2z(1+ht)\right]}{2z}+\frac{1}{8z^2}\right] \no\\
 &=& \frac{1}{\D(h)}
 \left[1-\frac{\cos2z}{2z}{{\cal D}(z)}
 +\frac{1}{8z^2}\right],
\ee
 where 
\be
 {\cal D}(z) = \int_{-\infty}^{\infty} dt 
 \frac{\cos (2z ht)}{\cosh^2 (2t)}
 = \frac{\pi h z/2}{\sinh (\pi h z/2)}.
\ee
 In a similar way, by using 
\be
 K(z_1,z_2) \sim \frac{\sin (z_1-z_2)}{z_1-z_2}-\frac{\cos (z_1+z_2)}{z_1+z_2},
\ee
 we obtain for the TLCF
\be
 \langle\langle\rho(\e_1)\rho(\e_2)\rangle\rangle 
 &\sim& -\frac{1}{\D^2(h)}
 \left[
 \frac{\sin (z_1-z_2)}{z_1-z_2}{\cal D}\left(\frac{z_1-z_2}{2}\right)
 -\frac{\cos (z_1+z_2)}{z_1+z_2}{\cal D}\left(\frac{z_1+z_2}{2}\right)
 \right]^2.
\ee
 As expected these results are in complete agreement with 
 the findings of the previous section by setting $h=1/\pi g$.
 We note that the function ${\cal D}(z)$ in Eq.(\ref{sdet}) 
 can be written in a spectral determinant form
\be
 {\cal D}(z) = \prod_{n=1}^{\infty}\frac{n^2}{n^2+h^2z^2/4}. \label{sdet2}
\ee
 We remark that
 the mapping proposed in this section 
 provides with explicit exact results for any value of $z$.
 Concerning the relation with the spectral properties
 at the Anderson transition we mention that in Ref.\cite{ant1} 
 it was shown analytically that the gchRMM 
 (and consequently our chRBM at $\alpha=1$) 
 reproduces all the features of critical statistics. 
 We do not repeat here this discussion and refer to it for details 
 and to the next section for numerical verification.

 Once we have proposed exact relations for the spectral correlations
 at the special value $\alpha=1$, 
 it is worthwhile to ask whether an analogous result can be extended 
 for the rest of $\alpha$s. 
 Unfortunately a mapping as the one above reported for $\alpha =1$
 cannot be extended to other values of $\alpha$. 
 We attempt a different strategy. 
 The DoS (\ref{dos1}) and the TLCF (\ref{tlcf1}) 
 were  explicitly  evaluated in Ref.\cite{ant1} 
 in the large conductance $g=1/\pi h \gg 1$ limit. 
 The resulting expressions, valid up to $1/g^2$ corrections, 
 were derived keeping the combination $z/g$ fixed. 
 This amounts to neglect $z^k/g^{k+2}$ terms as compared to $z^{k}/g^{k}$. 
 We remark that this expansion coincides with 
 the supersymmetry calculation in the AA domain but not 
 in the KM domain where $1/g^2$ corrections were kept.
 The results of Ref.\cite{ant1} (Eqs.(61) and (69) there)
 can be rewritten in the following way,
\be
 \rho_1(z) &=& \frac{\pi z}{2}(J_0^2(z)+J_1^2(z))
 +\frac{\pi}{2}J_0(z)J_1(z)({\cal D}(z)-1), \label{dos} \\
 \rho_2(x,y) &=& 
 \frac{1}{2}\mbox{Re}\sum_{q\ne 0}\left(\Pi^2_{+}+\Pi^2_{-}\right) 
 +\frac{{\cal D}_{+}^2-1}{2(x+y)^2}
 +\frac{{\cal D}_{-}^2-1}{2(x-y)^2} \no\\
 & & -\biggl\{
 \frac{\pi\sqrt{xy}}{2(x+y)}{\cal D}_+
 \left[J_1(x)J_0(y)+J_0(x)J_1(y)\right]
 +\frac{\pi\sqrt{xy}}{2(x-y)}{\cal D}_-
 \left[J_1(x)J_0(y)-J_0(x)J_1(y)\right]
 \biggr\}^2,  \label{conj2}
\ee
 where ${\cal D}_{\pm}={\cal D}((x\pm y)/2)$ and 
 $\Pi_{\pm}^2=\Pi^2(q,(x\pm y)/2)$ with $\cal D$ and $\Pi$ 
 given by Eqs.(\ref{sdet1}) and (\ref{pa}) 
 at $\alpha=1$ respectively.
 These expressions are consistent with Eqs.(\ref{dosaa}) and (\ref{2ptaa1}).
 For $\alpha=1$, 
 using the relation (\ref{pa1}), the first line of 
 the r.h.s of Eq.(\ref{conj2}) vanishes.
 The reason of keeping the expression of the first line 
 is that we want to separate purely perturbative contributions
 involving the propagator $\Pi$ from the nonperturbative ones.

 Since the dependence in $\alpha$ above is only through 
 ${\cal D}$ and $\Pi$,
 we speculate that for general $\alpha$ 
 the expressions for the DoS and the TLCF 
 of our chRBM (in the $g\gg 1$ limit) are given by 
 the above expressions with ${\cal D}$ and $\Pi$ 
 replaced by Eqs.(\ref{sdet}) and (\ref{pa}) respectively. 
 For the DoS, this speculation is supported by the supersymmetric calculation
 (which is valid for arbitrary $\alpha$) 
 since Eq.(\ref{dos}) is consistent with Eq.(\ref{dosaa}).
 For the TLCF the situation is less clear. 
 The supersymmetry result for the TLCF at $\alpha \neq 1$ has 
 a complicated form (\ref{2ptaa}) due to the nontrivial mixing 
 of perturbative and nonperturbative contributions.
 We remark that for the special value $\alpha=1$
 it was greatly simplified to Eq.(\ref{2ptaa1}) by 
 using nontrivial relations of the propagator 
 and the spectral determinant as demonstrated in Sec.\ref{alpha1}.
 We could not manage to recast (\ref{2ptaa}) as in Eq.(\ref{conj2}).  
 We thus conclude that Eq.(\ref{conj2}) for arbitrary $\alpha$
 is just the simplest alternative among several possible options.

\subsection{Numerical results}
\label{num}

\begin{figure}
\includegraphics[width=0.8\columnwidth,clip]{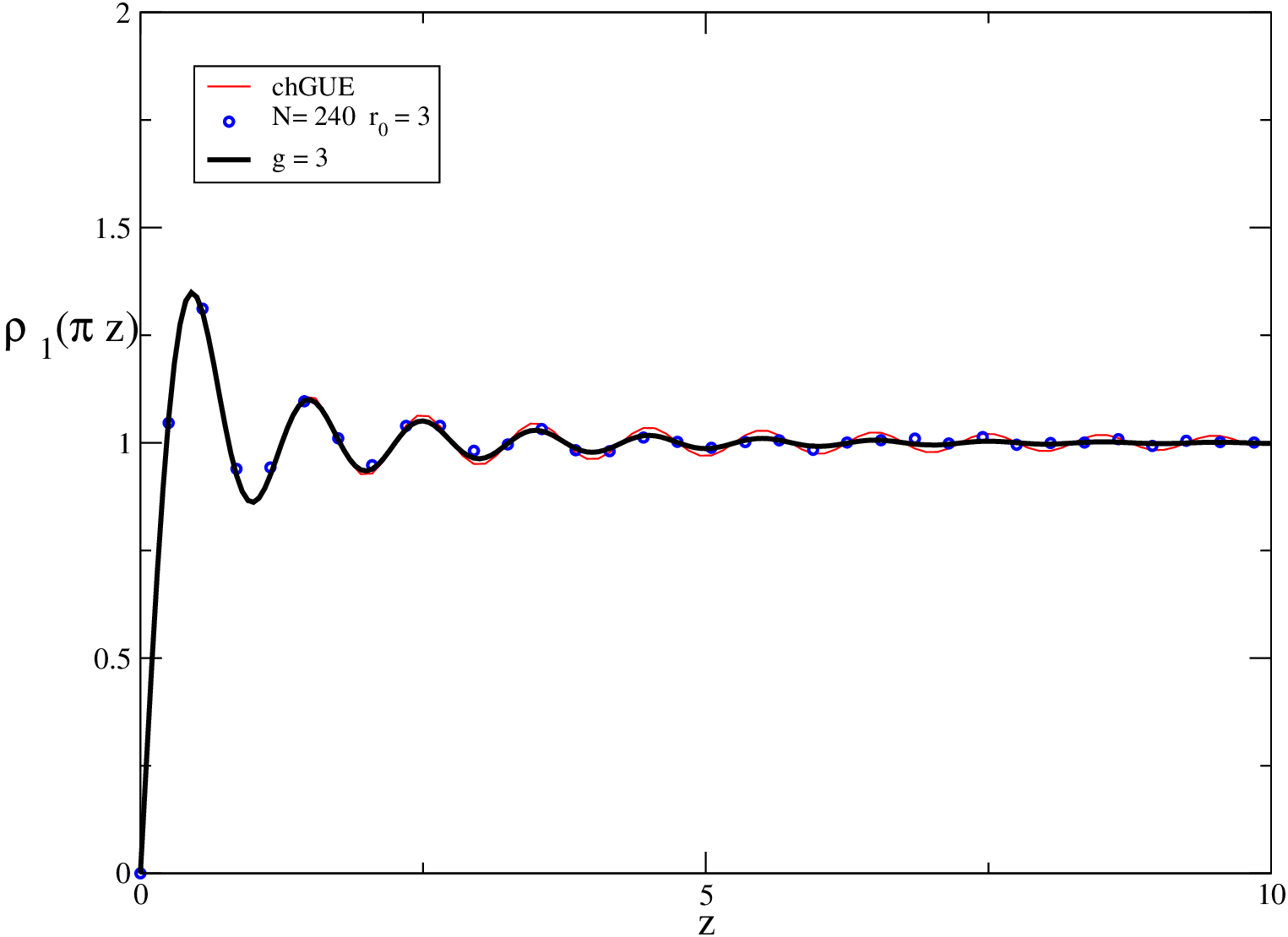}
\caption{DoS $\rho_1(z)$. 
 The solid line is the analytical prediction 
 of the chRBM (\ref{dos}) at a given $g$.
 The symbols represent the numerical results for 
 the indicated volume $N$ and bandwidth $r_0$, and 
 the thin line stands for the result at ergodic limit $g\to\infty$ 
 (chiral Gaussian unitary ensemble (chGUE)).  
 As observed the agreement with the analytical results is impressive.}  
\label{fig1} 
\end{figure}  

 We now compare the analytical predictions of previous sections 
 with the results of direct numerical diagonalization of the chRBM 
 in the spectral region close to the origin. 
 Since we are mainly interested in the spectral correlations 
 beyond the Thouless energy attention is focused on long range correlators 
 as the number variance 
 $\Sigma^{2}(L)=\langle L\rangle^2-\langle L^2\rangle
 =\int_{0}^{\pi L}dz\rho_1(z)-\int_{0}^{\pi L}dxdyR(x,y)$.  
 We recall that the number variance $\Sigma^{2}(L)$ measures
 the stiffness of the spectrum. 
 In the metallic regime, for eigenvalues separated less 
 than the Thouless energy, 
 fluctuations are small and $\Sigma^{2}(L)\sim\log(L)$ for $L\gg 1$. 
 Beyond the Thouless energy spectral fluctuations get stronger and
 $\Sigma^{2}(L)\sim L^{d/2}$, where $d$ is the dimensionality of the space.
 For disorder strong enough eigenvalues are
 uncorrelated (Poisson statistics) and $\Sigma^{2}(L)=L$. 
 At the Anderson transition, 
 the number variance is asymptotically proportional to 
 $\chi L$ ($\chi < 1$) \cite{chi}. 
 The spectral fluctuations are studied by direct diagonalization 
 of the chRBM for different sizes ranging from $N=2n=240$ to $N=2n=1200$. 
 The eigenvalues thus obtained are unfolded with respect to 
 the mean DoS.   
 The number of different realizations of disorder is chosen 
 such that for each $N$ the total number of eigenvalue be 
 at least $2 \times 10^5$.
 In order to reduce unwanted finite size effects we have utilized 
 a ``periodic boundary'' version of Eq.(\ref{rb}) as in Ref.\cite{prbm2}.

 We first study the spectral fluctuations at the critical value $\alpha=1$.
 In this case, since the spectral determinant can 
 be explicitly evaluated, both the DoS (\ref{dos}) 
 and the TLCF (\ref{conj2}) have a simple form.
 It is straightforward to show analytically \cite{ant1} that 
 $\Sigma^{2}(L)\sim L/4\pi r_0$ for $L \gg 1$. 
 In Fig.\ref{fig1} we show the numerical DoS versus 
 the analytical prediction (\ref{dos}). 
 As observed, the oscillations of the DoS are damped 
 with respect to the ergodic regime. 
 We stress that, since $g = r_0$, 
 the comparison between analytical and numerical is parameter free.

\begin{figure}
\includegraphics[width=0.8\columnwidth,clip]{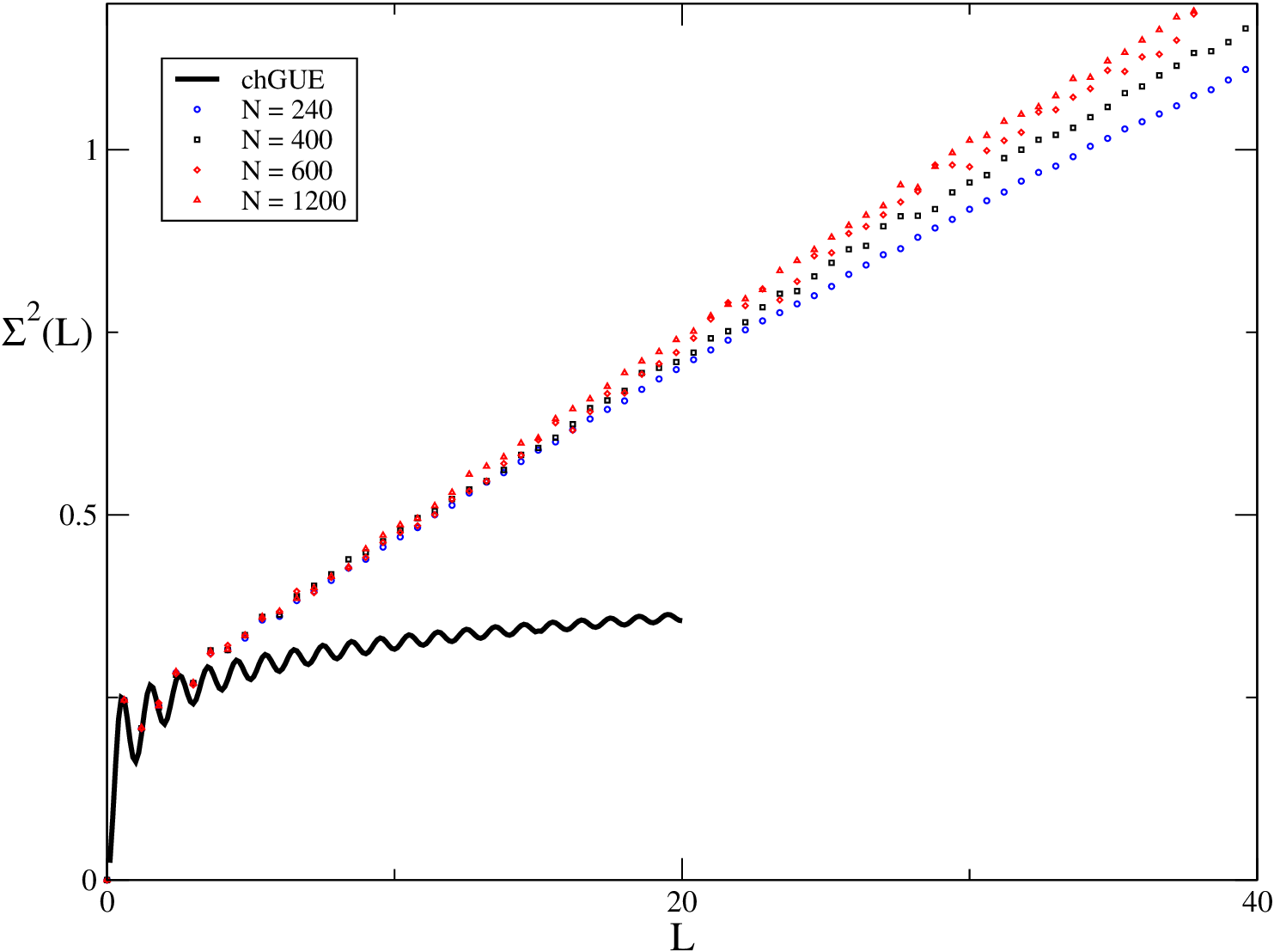}
\caption{Number variance $\Sigma^2(L)$ close to the origin of the spectrum. 
 The symbols represent the numerical results 
 for the indicated volume $N$ and bandwidth $r_0=2$.
 As observed, the spectrum is scale invariant. 
 Differences for large distances are due to finite size effects.}  
\label{fig2}
\end{figure}

\begin{figure}
\includegraphics[width=0.8\columnwidth,clip]{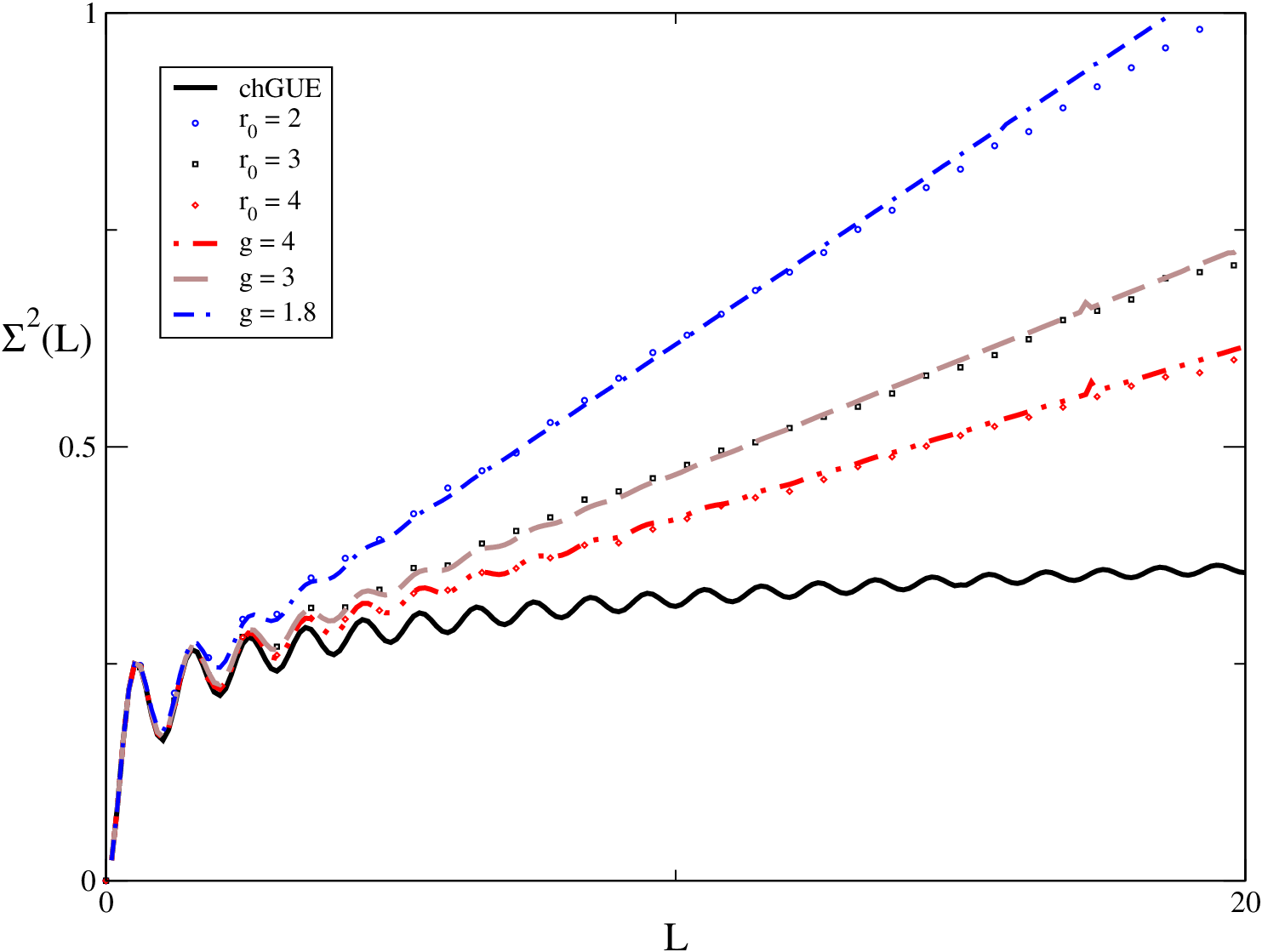}
\caption{Number variance $\Sigma^2(L)$ 
 close to the origin of the spectrum for $\alpha=1$. 
 The solid lines are the analytical predictions 
 for the chRBM Eqs.(\ref{dos}) and (\ref{conj2}), 
 and the symbols represent the numerical results for the
 indicated bandwidth $r_0$ and $N = 400$. 
 As observed, the agreement with  
 the analytical results is remarkable. 
 The value of $g$ chosen corresponds with the analytical prediction 
 except for $r_0=2$ where the best fitting is for $g=1.8$ instead of $g = 2$. 
 This deviation is expected since 
 the analytical predictions are valid only in the $g \gg 1$ limit.}  
\label{fig3} 
\end{figure}

 One of the signatures of an Anderson transition is
 the independence of $g$ on the system size. 
 In our case, analytically it is also predicted that 
 $g= r_0$ ($r_0 \gg 1$) is scale invariant.
 In Fig.\ref{fig2} the number variance
 for $r_0 = 3$ and $\alpha=1$ is plotted for different volumes $N$. 
 As expected the spectrum is scale invariant as at the Anderson transition.
 In Fig.\ref{fig3} we test our analytical findings  
 by comparing them with the numerical number variance at the origin 
 for different matrix sizes and bandwidth $r_0 = 3$. 
 As observed, the theoretical predictions are fully confirmed.
 The deviations appearing after $L \sim 20$ are
 a well understood finite size effect.  

\begin{figure}
\includegraphics[width=0.99\columnwidth,clip]{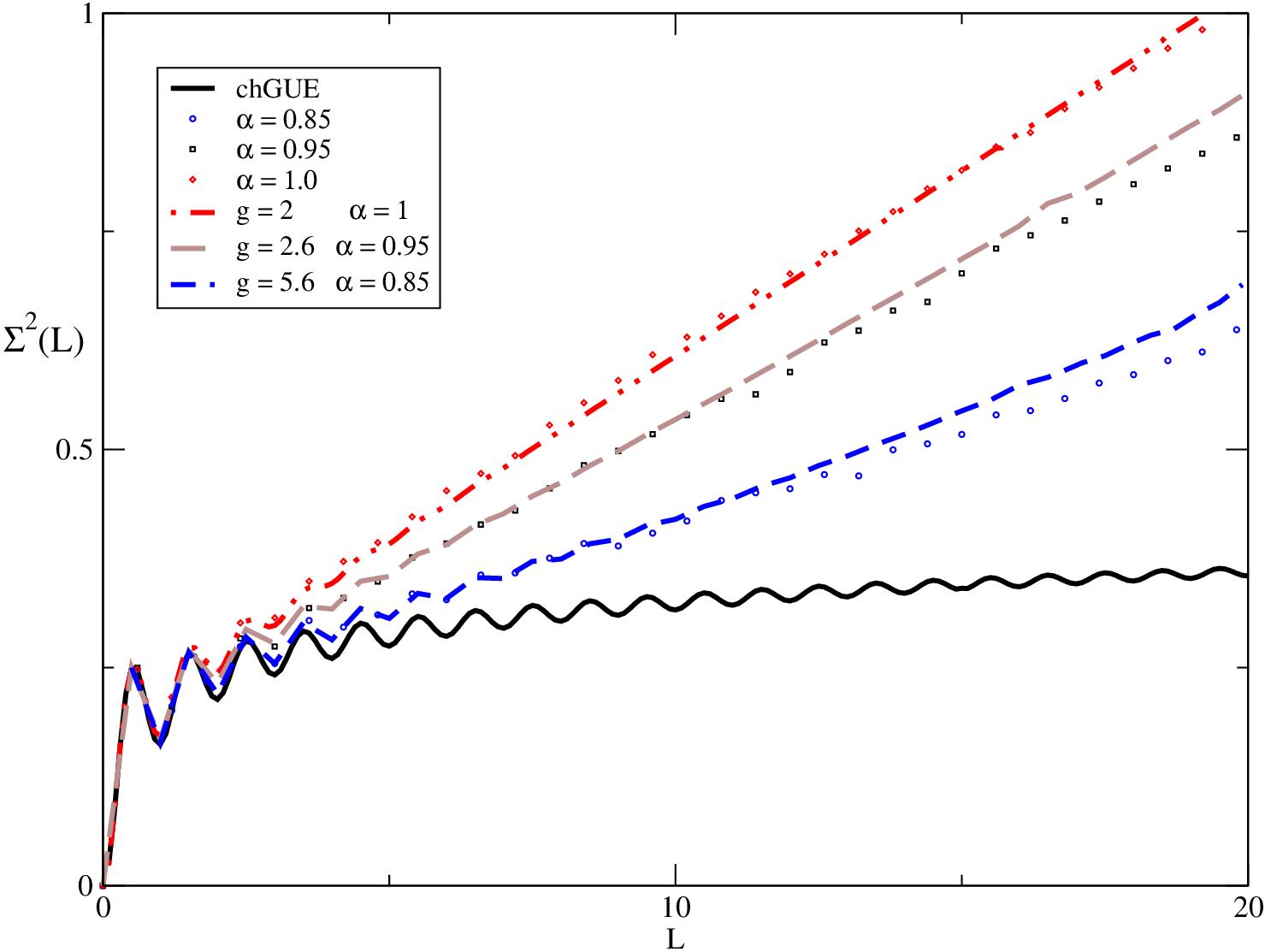}
\caption{Number variance $\Sigma^2(L)$ close to the 
 origin of the spectrum in the delocalized regime $\alpha \leq 1$. 
 The symbols correspond with the numerical simulation. 
 The lines correspond with the analytical number
 variance with $R(x,y)$ given by the conjecture Eq.(\ref{conj2}).
 In all cases $N$ in set to $N = 400$ and $r_0 = 2$. 
 The value of $g$ chosen is the best fit to the numerical results. 
 We checked that it is close ($\pm 5\%$) to the analytical prediction, 
 $g\sim r_0^{2\alpha-1} N^{2-2\alpha}$ \cite{prbm}.}
\label{fig4} 
\end{figure}

 We now study the spectral correlations for non critical values of $\alpha$. 
 For $1/2 <\alpha< 1$, according to our conjecture Eq.(\ref{conj2}),
 the asymptotic value of the TLCF is controlled by the power-law tail 
 of the perturbative part. 
 Thus it is expected that for $L \gg 1$, 
 $\Sigma^{2}(L) \sim L^{1/{2\alpha -1}}$. 
 The dimensionless conductance $g \sim r_0^{2\alpha -1}N^{2-2\alpha}$ 
 increases as the system size does. 
 In Fig.\ref{fig4} we show the numerical number variance 
 for different $\alpha$s versus the conjecture Eq.(\ref{conj2}). 
 The numerical number variance follows the expected asymptotic 
 power-law behavior in fair agreement with the conjectured result. 
 This evidence is not conclusive since
 different conjectures with the same asymptotic limits
 also may be in good agreement with the numerical results. 
 Clearly further work is needed to settle this issue. 

\section{Applications: the QCD vacuum as a disordered medium}
\label{app}

 In this section we discuss applications of 
 our results in the context of QCD.
 Generally, for $\alpha> 3/2$ corresponding to 
 normal diffusion, our model should describe the leading finite $g$ 
 corrections to the spectral correlations of any disorder system 
 with short range disorder and chiral symmetry. 
 The case of $1/2<\alpha\leq 3/2$ corresponds with 
 chiral disordered systems with long range hopping. 
 This is typical of systems driven 
 by dipole interactions ($1/r^3$) \cite{levitov}.
 We would also like to mention that for $\alpha = 1$ our results
 provide a phenomenological description of the properties of  
 systems with chiral symmetry at the Anderson transition.  
 Before going into details a word of caution is in order. 
 As mentioned in the introduction, 
 the spectral correlations of chiral systems,
 unlike for the standard universality class, are highly 
 dependent on microscopical details of the model 
 as the exact form of the disordered potential. 
 Thus the applications below reported must be considered 
 as educated guesses among a broad class of systems 
 in which our findings may be relevant. 

 In the infrared limit the eigenvalue correlations of the QCD Dirac operator
 do not depend on the dynamical details of the QCD Lagrangian
 but only on the global symmetries of the QCD partition function \cite{SV}. 
 Thus random matrices with the correct chiral symmetry
 of QCD (termed chiral random matrices) \cite{V} accurately describe 
 the spectral properties of the QCD Dirac operator up to some scale 
 known as the Thouless energy. 
 For larger energy differences, dynamic features become important and 
 the standard random matrix model ceases in principle to be applicable.
 However, in a recent work \cite{ant7}, it has been reported that 
 the spectral correlations of the QCD Dirac operator 
 in a background of instantons are accurately described 
 by a chiral RBM with $\alpha=3/4$ up to scales 
 well beyond the Thouless energy. 
 The reason for such exponent is indeed related to 
 the dipole-like interaction dominating 
 the interaction between quark zero modes and instantons. 
 The matrix elements $T_{\rm IA}$ (which physically describe 
 the amplitude of probability for a quark to hop from 
 an instanton to an antiinstanton) 
 of the QCD Dirac operator in a basis of chiral zero modes decay 
 as $T_{\rm AI} \sim 1/R_{\rm IA}^3$ 
 (dipole-like interaction) where $R_{\rm IA}$ 
 is the instanton-antiinstanton distance in a 4D space. 
 It was shown in Ref.\cite{levitov} that the spectral properties 
 of systems with power-law hopping are similar in different dimensions 
 provided that the decay exponent matches the space dimension. 
 Thus a decay $T_{\rm AI} \sim 1/R_{\rm IA}^3$ in 4D is equivalent to 
 $T_{\rm AI} \sim 1/R_{\rm IA}^{3/4}$ in 1D. 
 Furthermore, with this value of $\alpha$, 
 $g$ scales at the bulk as a 4D conductor 
 ($g \sim \sqrt{n}$)
 which is the expected result from chiral perturbation theory \cite{GL} 
 and lattice simulations \cite{lattice}. 
 The range of applicability of this chRBM was found 
 not to be restricted to the above mentioned zero temperature case.
 As usual in field theory, temperature is introduced
 by compactifying one of the spatial dimensions. 
 Thus in Euclidean QCD the effect of temperature is to reduce 
 the effective dimensionality of the system to three. 
 Now since the effective dimension of the space matches 
 the power-law decay of the QCD Dirac operator ($\sim 1/R^3$)
 one expects, according to Refs.\cite{PS,levitov}, 
 multifractal wavefunctions typical of a metal-insulator transition.
 As mentioned previously, this situation corresponds with 
 the case $\alpha=1$ in our chRBM.    
 The above findings suggest that, in case that the restoration of 
 the chiral symmetry expected at finite temperature were dominated 
 by instantons, the physical mechanism leading 
 to the quark-gluon plasma state of matter could be similar  
 to an Anderson transition driven by dipole interactions.

\section{Conclusion}
 We have studied the spectral properties of a disordered 
 system with chiral symmetry and long range hopping.
 By mapping the problem to a supersymmetric nonlinear $\sigma$ model 
 we have obtained 
 explicit expression for both the DoS and TLCF in different domains.
 It has been observed that, as in the nonchiral case, 
 if the power-law decay matches the dimensionality of the space the 
 spectral correlations are similar to the ones at the Anderson transition 
 and are described by critical statistics.
 Based on the relation with a gchRMM we have put forward 
 an exact formula for a specific value ($\alpha=1$) of the power-law decay 
 and we have also speculated that a similar relation should
 hold for the rest of $\alpha$s.
 Finally we have argued that in the context of QCD, 
 our model may be utilized to describe the spectral correlations of 
 the QCD Dirac operator beyond the Thouless energy 
 at zero and at finite temperature. 

 A.M.G. was supported by the EU network 
 ``Mathematical aspects of quantum chaos''.
 K.T was supported by SFB/Transregio 12.



\begin{thebibliography}{99}

\bibitem{WD}
 E.P. Wigner, Ann. Math. {\bf 53}, 36 (1951);
 F.J. Dyson, J. Math. Phys. {\bf 3}, 140, 157, 166 (1962).

\bibitem{SV}
 E.V. Shuryak and J.J.M. Verbaarschot, Nucl. Phys. {\bf A560}, 306 (1993).

\bibitem{SN}
 K. Slevin and T. Nagao, Phys. Rev. Lett. {\bf 70}, 635 (1993).

\bibitem{zirn} 
 M.R. Zirnbauer, J. Math. Phys. {\bf 37}, 4986 (1996).

\bibitem{AZ}
 A. Altland and M.R. Zirnbauer, Phys. Rev. B {\bf 55}, 1142 (1997).

\bibitem{V}
 J.J.M Verbaarschot, Phys. Rev. Lett. {\bf 72}, 2531 (1994).

\bibitem{VO}
 J.C. Osborn and J.J.M. Verbaarschot, Phys. Rev. Lett. {\bf 81}, 268 (1998); 
 Nucl. Phys. {\bf B525}, 738 (1998).

\bibitem{rfm} 
 A. Altland and B.D. Simons, Nucl. Phys. {\bf B562}, 445 (1999). 

\bibitem{GC}
 V. Gurarie and J.T. Chalker, Phys. Rev. Lett. {\bf 89}, 136801 (2002).

\bibitem{dwv}
 A. Altland, Phys. Rev. B {\bf 65}, 104525 (2002);
 W.A. Atkinson, P.J. Hirschfeld, A.H. Macdonald, and K. Ziegler, 
 Phys. Rev. Lett. {\bf 85}, 3926 (2000).

\bibitem{wire}
 A. Altland and R. Merkt, Nucl. Phys. {\bf B607}, 511 (2001);
 P. Brouwer, C. Mudry, and A. Furusaki, Phys. Rev. Lett. {\bf 84}, 2913 (2000). 

\bibitem{gade} 
 R. Gade, Nucl. Phys. B {\bf 398}, 499 (1993).

\bibitem{PL}
 C. P\'epin and P.A. Lee, Phys. Rev. Lett. {\bf 81}, 2779 (1998).

\bibitem{furu}
 A. Furusaki, Phys. Rev. Lett. {\bf 82}, 604 (1999).

\bibitem{AA}
 A.V. Andreev and B.L. Altshuler, Phys. Rev. Lett. {\bf 75}, 902 (1995);
 A.V. Andreev, B.D. Simons and B.L. Altshuler, 
 J. Math. Phys. {\bf 37}, 4968 (1996).

\bibitem{efetov}  
 K.B. Efetov, Adv. Phys. {\bf 32}, 53 (1983); 
 {\it Supersymmetry in Disorder and Chaos} 
 (Cambridge University Press, Cambridge, 1997).

\bibitem{wegner} 
 F. Wegner, Z. Phys. B {\bf 36}, 209 (1980).

\bibitem{janssen} 
 M. Janssen, Phys. Rep. {\bf 295}, 1 (1998). 

\bibitem{KMu} 
 V.E. Kravtsov and K.A. Muttalib, Phys. Rev. Lett. {\bf 79}, 1913 (1997).

\bibitem{SSSLS}
 B.I. Shklovskii, B. Shapiro, B.R. Sears, P. Lambrianides, and H.B. Shore, 
 Phys. Rev. B {\bf 47}, 11487 (1993).

\bibitem{chi} 
 B.L. Altshuler, I.K. Zharekeshev, S.A. Kotochigova, and B.I. Shklovskii, 
 Zh. Eksp. Teor. Fiz. {\bf 94}, 343 (1988) 
 [Sov. Phys. JETP {\bf 67}, 625 (1988)].

\bibitem{gRMM}
 M. Moshe, H. Neuberger, and B. Shapiro, 
 Phys. Rev. Lett. {\bf 73}, 1497 (1994); 
 K.A. Muttalib, Y. Chen, M.E.H. Ismail, and V.N. Nicopoulos,
 Phys. Rev. Lett. {\bf 71}, 471 (1993). 

\bibitem{prbm} 
 A.D. Mirlin, Y.V. Fyodorov, F.-M. Dittes, J. Quezada,
 and T.H. Seligman, Phys. Rev. E {\bf 54}, 3221 (1996).

\bibitem{levitov}
 L.S. Levitov, Phys. Rev. Lett. {\bf 64}, 547 (1990).

\bibitem{AL} 
 B.L. Altshuler and L.S. Levitov, Phys. Rep. {\bf 288}, 487 (1997).

\bibitem{ant5}
 A.M. Garcia-Garcia, cond-mat/0309445 (2003).

\bibitem{anderson}
 P.W. Anderson, Phys. Rev. {\bf 109}, 1492 (1958).

\bibitem{YO}
 G. Yeung and Y. Oono, Europhys. Lett. {\bf 4}, 1061 (1987).

\bibitem{YU} 
 C.C. Yu, Phys. Rev. Lett. {\bf 63}, 1160 (1989).

\bibitem{MK}
 R. Metzler and J. Klafter, Phys. Rep. {\bf 339}, 1 (2000).

\bibitem{prbm2}
 F. Evers and A.D. Mirlin, Phys. Rev. Lett. {\bf 84}, 3690 (2000);
 E. Cuevas, M. Ortuno et.al. Phys. Rev. Lett. {\bf 88}, 016401 (2002).

\bibitem{ktaka} K. Takahashi, cond-mat/0403284 (2004).

\bibitem{MI}
 N. Mae and S. Iida, J. Phys. A {\bf 36}, 999 (2003).

\bibitem{KM} 
 V.E. Kravtsov and A.D. Mirlin, JETP Lett. {\bf 60}, 656 (1994).

\bibitem{ant1}
 A.M. Garcia-Garcia and J.J.M. Verbaarschot, 
 Nucl. Phys. {\bf B586}, 668 (2000).

\bibitem{VZ} 
 J.J.M. Verbaarschot and I. Zahed, Phys. Rev. Lett. {\bf 70}, 3852 (1993).

\bibitem{AST}
 A.V. Andreev, B.D. Simons, and N. Taniguchi, 
 Nucl. Phys. B {\bf 432}, 487 (1994).

\bibitem{KT}
 V.E. Kravtsov and A.M. Tsvelik, Phys. Rev. B {\bf 62}, 9888 (2000). 

\bibitem{ant3} 
 A. M. Garcia-Garcia and J.J.M. Verbaarshot,
 Phys. Rev. E {\bf 67}, 046104 (2003).

\bibitem{CS}
 F. Calogero, J. Math. Phys. {\bf 10}, 2191 (1969);
 B. Sutherland, J. Math. Phys. {\bf 12}, 246 (1971).

\bibitem{mirlin}
 A.D Mirlin, Phys. Rep. {\bf 326}, 259 (2000).

\bibitem{ant7}
 A.M. Garcia-Garcia and J.C. Osborn, hep-th/0312146 (2003).

\bibitem{GL}
 J. Gasser and H. Leutwyler, 
 Phys. Lett. {\bf 188B}, 477 (1987); Nucl. Phys. {\bf B307},763 (1988).

\bibitem{lattice}
 M.E. Berbenni-Bitsch, M. G\"ockeler, T. Guhr,
 A.D. Jackson, J.-Z. Ma, S. Meyer, A. Sch\"afer,
 H.A. Weidenm\"uller, T. Wettig, and T. Wilke,
 Phys. Lett. B {\bf 438}, 14 (1998);
 M. Gockeler, H. Hehl, P.E.L. Rakow, A. Schafer, and T. Wettig, 
 Phys. Rev. D {\bf 59}, 094503 (1999).

\bibitem{PS}
 A. Parshin and H.R. Schober, Phys. Rev. B {\bf 57}, 10232 (1998).

\end{thebibliography}
\end{document}